\documentclass[conference,a4paper]{IEEEtran}

\usepackage{balance}
\usepackage{subfigure}
\usepackage{amssymb}
\usepackage{color}

\usepackage[utf8]{inputenc} 
\usepackage[T1]{fontenc}
\usepackage{url}              % provides \url{...}
\usepackage{cite}             % improves presentation of citations

\usepackage[cmex10]{amsmath}  % Use the [cmex10] option to ensure complicance
\usepackage{mleftright}       % fix to wrong spacing of \left-,
\mleftright                   % \middle- \right-commands 

\usepackage{graphicx}         % provides \includegraphics{...} to
\usepackage{booktabs}         % fixes poor spacing in tables and
\newtheorem{lemma}{Lemma}
\newtheorem{remark}{Remark}
\newtheorem{corollary}{Corollary}
\newtheorem{theorem}{Theorem}
\newtheorem{assumption}{Assumption}

\newtheorem{propo}{Proposition}

\usepackage{xspace}
\usepackage{bbm}
\usepackage{mathrsfs}
\newcommand{\Dc}{\mathcal{D}}

\newcommand{\Fc}{\mathcal{F}}

\newcommand{\Hc}{\mathcal{H}}

\newcommand{\Pc}{\mathcal{P}}

\newcommand{\Rc}{\mathcal{R}}

\newcommand{\Pv}{{\bf P}}

\newcommand{\kh}{{\hat{k}}}

\def\a{\alpha}

\def\l{\lambda}

\newcommand{\Poisson}{\mathrm{Pois}}

\newcommand{\U}{\mathrm{Unif}}

\def\textiid{i.i.d.\@\xspace}
\newcommand\iid{\ifmmode\text{ i.i.d. } \else \textiid \fi}

\newcommand{\Real}{\mathbb{R}}

\newcommand{\ind}{\boldsymbol{1}}

\begin{document}

\title{Sample-and-Forward: Communication-Efficient Control of the False Discovery Rate in Networks} 

%%%%%%
% \author{%
%   \IEEEauthorblockN{Anonymous Authors}
%   \IEEEauthorblockA{%
%     Please do NOT provide authors' names and affiliations\\
%     in the paper submitted for review, but keep this placeholder.\\
%     ISIT23 follows a \textbf{double-blind reviewing policy}.}
% }

\author{\IEEEauthorblockN{Mehrdad Pournaderi and Yu Xiang}
\IEEEauthorblockA{\textit{Department of Electrical and Computer Engineering}\\
\textit{University of Utah}\\
Salt Lake City, UT 84112, USA \\
\{m.pournaderi,\,yu.xiang\}@utah.edu}
}

\maketitle

%%%%%
%% Abstract: 
%% If your paper is eligible for the student paper award, please add
%% the comment "THIS PAPER IS ELIGIBLE FOR THE STUDENT PAPER
%% AWARD." as a first line in the abstract. 
%% For the final version of the accepted paper, please do not forget
%% to remove this comment!
%%
\begin{abstract}
This work concerns controlling the false discovery rate (FDR) in networks under communication constraints. We present sample-and-forward, a flexible and communication-efficient version of the Benjamini-Hochberg (BH) procedure for multihop networks with general topologies. Our method evidences that the nodes in a network do not need to communicate p-values to each other to achieve a decent statistical power under the global FDR control constraint. Consider a network with a total of $m$ p-values, our method consists of first sampling the (empirical) CDF of the p-values at each node and then forwarding $\mathcal{O}(\log m)$ bits to its neighbors. Under the same assumptions as for the original BH procedure, our method has both the provable finite-sample FDR control as well as competitive empirical detection power, even with a few samples at each node. We provide an asymptotic analysis of power under a mixture model assumption on the p-values.
\end{abstract}

\section{Introduction}

% Our work is inspired by a distributed version of the FDR control problem~\cite{Ramdas2017b}, where agents in a network aggregate and exchange statistical test results in a collaborated manner in order to control the overall FDR.
\label{sec:intro}
%
% The multiple testing problem, motivated by applications in microarray, has been an active research area across various fields. Since the concept of the false discovery rate (FDR) control being formally introduced by Benjamini and Hochberg in~\cite{benjamini1995}, it has attracted great interest due to its wide applications in areas such as genomics and neuroimaging~\cite{genovese2002thresholding,abramovich2006adapting}, and many extensions have been investigated~\cite{benjamini2001control,efron2001empirical,storey2002direct,sarkar2002some,genovese2002operating,efron2012large}.
\IEEEPARstart{F}{or} large-scale networks where each node performs multiple hypothesis tests, it is of theoretical and practical interest to understand the amount of communication needed among the nodes to achieve a satisfactory testing performance. We follow the false discovery rate control criterion and aim to adapt the celebrated Benjamini-Hochberg (BH) procedure~\cite{benjamini1995} (also see~\cite{benjamini2001control,efron2001empirical,storey2002direct,sarkar2002some,genovese2002operating,efron2012large,genovese2002thresholding,abramovich2006adapting} for extensions) to the network settings \emph{with a limited communication budget in mind} (e.g., battery-powered sensors), where multiple hypothesis tests come in to play naturally when each node in the network is equipped with multiple sensors for various measurements (e.g., temperature, PM2.5, humidity, etc). It should be pointed out that our focus on decentralized inference with FDR control is fundamentally different from the existing literature on distributed detection and hypothesis testing formulations~\cite{tenney1981,tsitsiklis1984,viswanathan1997,blum1997} which focus on testing a single hypothesis. 

Our study is motivated by a recent distributed FDR control method called the Query-Test-Exchange (QuTE) algorithm~\cite{Ramdas2017b}, which requires each node to transmit its p-values to all of its neighbors. This method can handle \emph{multihop networks with general topologies}, however, the amount of communication needed to implement this algorithm can be a bottleneck for practical problems as the size of the network or the number of p-values per node grows. 
%There are two natural baselines that serve as two extreme cases. One requires no communication, where each node performs the BH procedure (with a corrected test size) on its own p-values, leading to a Bonferroni type test; on the other hand, the global BH at each node is achievable if unlimited communication is allowed. A natural question in a distributed network is whether it is possible to achieve FDR control with good power in a communication-efficient manner. 
On the other hand, the authors in~\cite{ermis2005adaptive,ray2007novel,ermis2009distributed,ray2011false} have studied a distributed sensor network setting by assuming a \emph{broadcast model}, where each sensor is allowed to broadcast its decision to the entire network. Under this model, they have proposed several (iterative) distributed BH procedures where each sensor broadcasts at most $1$-bit information (which is possible as each sensor has only one p-value in~\cite{ermis2005adaptive,ray2007novel,ray2011false}; or a transformation is needed in~\cite{ermis2009distributed} to combine multiple dependent p-values to a single scalar). Despite the low communication cost, this broadcast model is mainly suitable for small-scale sensor networks (e.g., see~\cite[Section VIII]{ermis2009distributed} for discussions), and the proposed iterative methods are not applicable for multihop networks with general topologies, which is the main focus of this work. In~\cite{pournaderi2022communication,pournaderi2022large}, the authors took an \emph{asymptotic} perspective and proposed a communication-efficient aggregation method for star networks to match the global performance by achieving the global BH threshold asymptotically. However, the analysis is made possible under a mixture model assumption and a fixed alternative (conditional) distribution throughout the nodes, which is different from the original setting for the BH procedure with no restrictions on the alternative distributions. It has been shown that it is possible to reduce the communication cost of the QuTE algorithm 
%by incorporating discrete FDR control methods
via quantization~\cite{xiang2019distributed,Ramdas2022}, however, these methods still require each node to transmit the quantized version of the p-values.

%In~\cite{pournaderi2022communication}, the authors took an \emph{asymptotic} perspective and proposed a communication-efficient aggregation method to match the global performance by achieving the global BH threshold asymptotically. However, the analysis is made possible under a mixture model assumption and a fixed alternative (conditional) distribution throughout the nodes, which is different from the original setting for the BH procedure with no restrictions on the alternative distributions; moreover, the finite sample performance can only be shown empirically without theoretical guarantees. A related attempt to reduce the communication cost of the QuTE algorithm has been made in~\cite{xiang2019distributed}, but it still requires transmitting the quantized version of all the p-values in a network. 

 In this work, we 
 %focus on multihop networks with general topologies, and 
 propose \emph{sample-and-forward}, a sampling-based version of the BH procedure for networks, with \emph{finite-sample} provable FDR control. In our method, each node 
 samples the (empirical) CDF of p-values (see~\eqref{eq:pseudo}) 
 %computed based on its own observed p-values and only needs to send \emph{$(M+1)\log(m)$ bits} to a center node, where $M$ is the number of samples each node takes;
 and 
 only needs to send $\mathcal{O}(\log m)$ \emph{bits} to its neighboring nodes.
 %This is in contrast with sending $\mathcal{O}(m)$ \emph{real-valued} p-values required in the QuTE method~\cite{Ramdas2017b} (or $\mathcal{O}(m)$ bits for the quantized QuTE~\cite{Ramdas2022}).
 This is in contrast with sending $\mathcal{O}(m)$ real-valued p-values required in the QuTE method~\cite{Ramdas2017b}; $\mathcal{O}(m \log m)$ bits are needed for the quantized QuTE~\cite{Ramdas2022} to retain the full power of BH, while $\mathcal{O}(m)$ bits are sufficient for the FDR control.
% %, where $m^{i}$ denotes the number of p-values node~$i$ possesses.
 Our method is flexible in the sense that (a) it can be coupled with the QuTE algorithm to deal with arbitrary network topologies, and (b) the FDR control holds independent of the number of samples taken at each node. Our method shows robust and competitive statistical powers in a variety of simulation settings. 
\section{Multiple Testing and FDR Control}
\label{sec:background}

% \subsection{}
% \label{sec:FDR}

%Let $\mathsf{H_0}$ and $\mathsf{H_1}$ denote the null and alternative hypothesis, respectively.
%For test statistics $\{X_1,...,X_m\}$,

Consider testing the null hypotheses $\mathsf{H_{0,k}, 1 \leq k\leq m}$ according to the test statistics $X_k, 1 \leq k\leq m$. Let $P_k = \Psi_{\mathsf{H_{0,k}}}(X_k)$, $1\le k\le m$, denote the p-values computed for the test statistics, where $\Psi_{\mathsf{H_{0,k}}}=1-\mathsf{F}_{\mathsf{H_{0,k}}}$ and $\mathsf{F}_{\mathsf{H_{0,k}}}$ is the (hypothetical) CDF of $X_k$ under $\mathsf{H_{0,k}}$. Let $m_0$ denote the number of statistics generated according to their corresponding null hypotheses, i.e., $m_0$ of the null hypotheses are true. The p-values computed for these $m_0$ statistics are called \emph{null p-values} and we refer to the rest of them as \emph{non-nulls}. If the statistics are generated according to a continuous distribution functions, by the probability integral transform, the null p-values will have a uniform distribution over $[0,1]$.

The multiple testing problem concerns testing $m$ hypotheses while controlling a simultaneous measure of the type I error at some (prefixed) level $\alpha$. The two most popular approaches are the family-wise error rate (FWER) control and the false discovery rate (FDR) control. The FWER is more conservative (reject fewer hypotheses) as it controls the probability of making at least one false rejection, while FDR is more liberal by controlling the expected \emph{proportion} of false rejections among all rejections. Let $R$ and $V$ denote the total number of rejections and the number of false rejections, respectively. Then, FWER is defined as $\mathbb{P}(V>0)$ and it can be controlled at some target level $\alpha$ by rejecting $\Rc_{\text{Bonf}}=\{k: P_k \leq \alpha/m\}$. The threshold $\alpha/m$ is the so-called Bonferroni-corrected test size. On the other hand, FDR is defined as $\text{FDR} = \mathbb{E}\left(\frac{V}{R\vee 1}\right)$
% \begin{equation}
% 	\text{FDR} = \mathbb{E}\left(\frac{V}{R\vee 1}\right),
% \end{equation}
and the celebrated Benjamini-Hochberg (BH) procedure~\cite{benjamini1995} controls the $\text{FDR}$ at level $\a$ under the \emph{Positive Regression Dependence on a Subset} (PRDS) assumption \cite{benjamini2001control}.
% \begin{define}[PRDS Condition \cite{benjamini2001control,ramdas2019unified}]
The vector of p-values $\Pv=(P_{1},P_{2},\ldots, P_{m})$ is said to satisfy the PRDS condition if $\mathbb{P}\left(\Pv\in\Dc|P_i\leq t\right)$ is non-decreasing over $(0,1]$ for all non-decreasing $\Dc\subseteq[0,1]^m$ and $i\in\Hc_0:=\{ k:\mathsf{H_{0,k}}\text{ is true}\}$~\cite{ramdas2019unified}.
% \end{define}

\smallskip
\begin{assumption}\label{prds}
$\Pv$ satisfies the PRDS condition. The PRDS condition holds trivially if the p-values are independent~\cite{benjamini2001control}.
%The null p-values are mutually independent and independent of the non-null p-values.%\footnote{This assumption can be relaxed to the PRDS condition in \cite{benjamini2001control}.}
\end{assumption}
\smallskip

Let $(P_{(1)},P_{(2)},\ldots, P_{(m)})$ denote the ascending-ordered p-values. Using this notation, BH procedure rejects $\hat{k}$ smallest p-values, i.e., $\Rc_{\text{BH}}=\{k: P_k \leq P_{(\hat{k})}\}$, where 
$\kh = \max\big\{1\leq k\leq m: P_{(k)}\le \tau_k \big\}$
% \begin{equation*}
%     \kh = \max\big\{1\leq k\leq m: P_{(k)}\le \tau_k \big\}
% \end{equation*}
with $\tau_k=\a k/m$ 
and $\hat{k}=0$ if $P_{(k)} > \tau_k$ for all $k$. We adopt the notation $R_{\text{BH}}:=\hat{k}$ and $\tau_{\text{BH}}: = \tau_{\hat{k}}=\alpha R_{\text{BH}}/m$ for the rejecting index and rejecting threshold, respectively. The power of detection for a multiple testing procedure is defined as $\Pc={\mathbb{E}\,\text{TDP}}$ where $\text{TDP}=\frac{R-V}{m_1\vee 1}$ is the true discovery proportion, where $a\vee b:=\max\{a,b\}$. Define the function
 \begin{equation}
 \label{eq:pseudo}
     \mathcal{F}(t)=\frac{1}{m}\sum_{i=1}^m\ind\{P_i\leq t\}.
 \end{equation}
In the rest of this work, we refer to  $\mathcal{F}(t)$ as the \emph{pseudo-CDF} since the p-values are not assumed to be \iid. Observe that $\mathcal{F}(P_{(k)})=k/m$. Therefore $R_{\text{BH}}=\max\big\{0\leq k\leq m:  \mathcal{F}(P_{(k)})\geq P_{(k)}/\alpha \big\}$,
%\begin{align*}
   % R_{\text{BH}} %&= \max\big\{0\leq k\leq m: P_{(k)}\le \alpha \mathcal{F}(P_{(k)}) \big\}\\
       % &= \max\big\{0\leq k\leq m:  \mathcal{F}(P_{(k)})\geq P_{(k)}/\alpha \big\},
%\end{align*}
with the convention $P_{(0)}=0$. Similarly, an alternative representation for the BH threshold $\tau_\text{BH}$ is known \cite{genovese2004stochastic}, which can be adapted to our setting straightforwardly as $ \tau_{\text{BH}}=\sup\{t\geq 0:\mathcal{F}(t)= t/\alpha\}.$
%\begin{lemma}
% \begin{align*}
%     \label{rep}
% \end{align*}$
%\end{lemma}
% \begin{proof}
% The proof is trivial for $R_{\text{BH}} = m$ and $R_{\text{BH}}=0$.
% % \begin{equation*}
% %     \tau_{\text{BH}}=\alpha = \sup\{t\geq 0:\mathcal{F}(t)= t/\alpha\} . 
% % \end{equation*}
% % If $R_{\text{BH}}=0$, then 
% Now suppose $0<R_{\text{BH}}<m $. We first show that $\mathcal{F}(\tau_{\text{BH}}) = \tau_{\text{BH}}/\alpha$. By the definition of $R_{\text{BH}}$ we have $P_{(R_{\text{BH}})}\leq \tau_{\text{BH}} <P_{(R_{\text{BH}}+1)}$, and thus $\mathcal{F}(\tau_{\text{BH}})=\mathcal{F}(P_{(R_{\text{BH}})})=R_{\text{BH}}/m$. Also, we have $\tau_{\text{BH}}/\alpha=R_{\text{BH}}/m$ since $\tau_{\text{BH}}=\a R_{\text{BH}}/m$ by definition. Hence, $\mathcal{F}(\tau_{\text{BH}}) = \tau_{\text{BH}}/\alpha=R_{\text{BH}}/m$.

% Now we show that $\tau_{\text{BH}}$ is the largest value satisfying $\mathcal{F}(\tau_{\text{BH}}) = \tau_{\text{BH}}/\alpha$. Suppose for contradiction that $\mathcal{F}(t') = t'/\alpha$ for some $t'>\tau_{\text{BH}}=\a R_{\text{BH}}/m$. If $t'\geq P_{(R_{\text{BH}}+1)}$, then we get $R_{\text{BH}}=R_{\text{BH}}+1$, which is a contradiction. If $\a R_{\text{BH}}/m<t'<P_{(R_{\text{BH}}+1)}$, we get $\mathcal{F}(t')=R_{\text{BH}}/m<t'/\a$. This contradicts with our assumption that $\mathcal{F}(t') = t'/\alpha$, completing the proof.
% \end{proof}

\section{Sample-and-Forward for Star Networks}
\label{sec:quantized-FDR}
To set the stage for the general multihop network settings in the next section, we start with describing our method for {a star network, where one center node can send (or receive) information to (or from) $N$ other nodes. At each node $i\in \{1,...,N\}$, there are $m^{(i)}=m^{(i)}_0 + m^{(i)}_1$ p-values $\Pv^{(i)}=(P_1^{(i)},..., P_{m^{(i)}}^{(i)})$, where $m^{(i)}_0$ (or $m^{(i)}_1$) of them correspond to the true null (or alternative) hypotheses.}

There are two obvious ways to attain the global FDR control over the network. Baseline~I: performing the global BH procedure, i.e., pooling all the p-values in the network at the center node and performing the BH procedure; Baseline~II: performing local BH procedures with (node-wise) Bonferroni corrected test size $\alpha^{(i)} = \frac{m^{(i)}}{m}\alpha$. 
For Baseline~I, it is straightforward to observe that the same performance can be achieved by only communicating the p-values less than the test size $\alpha$ along with the number of p-values each node possesses. However, this is not communication-efficient as it still requires each node to communicate $\mathcal{O}(m)$ p-values. Regarding the second baseline, even though no communication is needed, the method suffers from a significant power loss in comparison to the pooled inference, especially as the number of nodes in the network grows. Our proposed algorithm achieves a detection power close to Baseline~I but with $\mathcal{O}(\log m)$ \emph{bits} communication cost for each node as we shall see in the section on numerical results. 
% \vspace{-1ex}

\subsection{Sample-and-Forward: Star Networks}\label{sec:alg}
For an overall targeted FDR level $\alpha$, our \emph{sample-and-forward BH method} consists of four main steps. Note that $\tau_{\text{BH}}\le \alpha$, thus it suffices to sample in $[0,\alpha]$.
\begin{itemize}\setlength\itemsep{0.06em}
	\item[(1)] {\bf Sample the pseudo-CDF}:  Each node~$i$ computes $\mathcal{F}_i(t)=\frac{1}{m^{(i)}}\sum_{j=1}^{m^{(i)}}\ind\{P_j^{(i)}\leq t\}$ and then take $M\geq 2$ samples from it on the $[0,\alpha]$ interval, at locations  $t_j:=\frac{j-1}{M-1}\alpha,\ 1\leq j\leq M$. Then each node $i$ sends (I) the samples (i.e., $\mathcal{F}_i(t_j),\ 1\le j\le M$) and (II) the number of p-values $m^{(i)}$, to the center node. 

 	\item[(2)] {\bf Approximate the pooled pseudo-CDF}: The center node computes $S_j=\frac{1}{m}\sum_{i=1}^N{m^{(i)}{\mathcal{F}}_i(t_j)},\ 1\leq j\leq M$ where $m=\sum_{i=1}^N{m^{(i)}}$.
	\item[(3)] {\bf Compute the global threshold}: The center node computes  $\hat\tau=t_I+\alpha\big(S_I-t_I/\alpha\big)$ where $I=\max\big\{j:  S_j\geq t_j/\alpha \big\}$ and broadcasts it to the other nodes.
	
	\item[(4)] {\bf Reject according to $\hat\tau$}: Upon receiving $\hat\tau$, each node $i$ rejects $\{1\leq j\leq m^{(i)}:P_j\leq \hat\tau\}$.
	
\end{itemize}

% \begin{remark}
% The BH procedure rejection threshold always lies inside $[0,\alpha]$. Thus it suffices to sample in this interval.
% \end{remark}

In the statement of the algorithm, we take a fixed number of samples $M$ at fixed locations $t_j,\ 1\leq j\leq M$ for each node to simplify the presentation. Nevertheless, the proof does not rely on this structure. In fact, each node (i) can take an arbitrary number of samples $M^{(i)}$ at any set of locations $\big\{Z_1^{(i)},Z_2^{(i)},\hdots,Z_{M^{(i)}}^{(i)}\big\}$. However, these locations need to be communicated to the center node in step (1). To avoid this communication cost, one can restrict the sampling positions to be equispaced. In this case, the center node can compute the positions based on the number of samples it receives from the local nodes. For the general framework, in step (2), the pooled pseudo-CDF is approximated as follows,
\begin{equation}
    \hat{\mathcal{F}}(t) =\frac{1}{m}\sum_{i=1}^N{m^{(i)}\hat{\mathcal{F}}_i(t)}, \label{eq:pooling}
\end{equation}
where $\hat{\mathcal{F}}_i(t)=0$ for $t<Z_1^{(i)}$, $\hat{\mathcal{F}}_i(t)=1$ for $t\ge 1$, and 
\begin{equation}
    \hat{\mathcal{F}}_i(t) = 
     \left\{
	\begin{array}{ll}
		\mathcal{F}_i(Z_j^{(i)}), \qquad\ \,  Z_j^{(i)}\leq t<Z_{j+1}^{(i)}\\
		\mathcal{F}_i(Z_{M^{(i)}}^{(i)}), \quad\ \ \, Z_{M^{(i)}}^{(i)}\leq t < 1
	\end{array}
\right.\  \label{stime}
\end{equation}
% \begin{equation}
%     \hat{\mathcal{F}}_i(t) = 
%      \left\{
% 	\begin{array}{ll}
% 		0,\qquad\qquad\qquad\quad\ \ \ \,\ t<Z_1^{(i)} \\
% 		\mathcal{F}_i(Z_j^{(i)}), \qquad\ \,  Z_j^{(i)}\leq t<Z_{j+1}^{(i)}\\
% 		\mathcal{F}_i(Z_{M^{(i)}}^{(i)}), \quad Z_{M^{(i)}}^{(i)}\leq t < 1\\
% 		1, \qquad\qquad\qquad\quad\ \ \ \ \,   t\geq 1
% 	\end{array}
% \right.\  \label{stime}
% \end{equation}
and $1\leq j<M^{(i)}$.
The general version of the formula given in step (3) (for $\hat\tau$) can be defined as
\begin{equation}
    \hat\tau=\sup\left\{t\geq 0:\hat{\mathcal{F}}(t)= t/\alpha\right\}.\label{eq:BH-th}
\end{equation}
To implement this formula, one can use the following procedure. Let $\{J_1,\hdots,J_M\}$ denote the set of jump locations of $\hat{\mathcal{F}}(t)$. Define $\tau'=\max\big\{J_i:  \hat{\mathcal{F}}(J_i)\geq J_i/\alpha \big\}$. The threshold $\hat\tau$ can be computed via $\hat\tau=\tau'+\alpha\big(\hat{\mathcal{F}}(\tau')-\tau'/\alpha\big)$ which is compatible with the formula given in step (3).

% \begin{figure}[!h]
% \centering
%   \includegraphics[scale=0.27]{curves.png}
%   \caption{The black solid intervals represent $h(t)$ and the red dotted intervals represent $\hat{h}(t)$. The area between the two curves is upper bounded by the sum of the shaded areas.}
%   \label{fig:curves}
% \end{figure}

% To gain some insight into the approximation (see Section~\ref{sec:quality} for a rigorous treatment), we define $h=\mathcal{F}_{[0,\alpha]}$ and $\hat{h}=\hat{\mathcal{F}}_{[0,\alpha]}$, and look into $\|h-\hat{h}\|_1$.
%  As illustrated in Figure~\ref{fig:curves}, $[0,\a]$ is divided into $M$ interval each with length $\a/M$, and each jump of the function $h(t)$ is of size $1/m$. To compute an upper bound for $\|h-\hat{h}\|_1$, we observe that the number of p-values within each interval determines the number of jumps in that interval. Since the area between the two curves can be upper bounded by the shaded rectangular regions, one can show that 
% \begin{align*}
% 	\int_{0}^\alpha \big|h(t)-\hat{h}(t)\big|\, d\mu(t)
% 	&\le \sum_{i=1}^M \frac{L_i}{m}\cdot\frac{\a}{M}\le \frac{\a}{M},
% \end{align*} 
% where $L_i$ denotes the number of p-values in the $i$-th interval. 

\subsection{FDR Control}
Let ${\mathcal{F}}(t) =\frac{1}{m}\sum_{i=1}^N{m^{(i)}{\mathcal{F}}_i(t)}$ denote the (pooled) pseudo-CDF of the p-values in the network. 
% \begin{lemma} \label{FDRlem}
% Let $\widetilde{\mathcal{F}}(t)$ denote any non-decreasing function such that $\widetilde{\mathcal{F}}(t)\leq {\mathcal{F}}(t)$ for all $t\in\Real$. Then, rejecting according to $\widetilde{\tau}=\sup\{t\geq 0:\widetilde{\mathcal{F}}(t)= t/\alpha\}$ controls the FDR.
% \end{lemma}
% \begin{proof}
% For each realization of p-values, there exist a $\Breve{\alpha}\leq \alpha$ such that $\mathcal{F}(\widetilde{\tau})=\widetilde{\tau}/\Breve{\alpha}$ since $\widetilde{\mathcal{F}}(t)\leq {\mathcal{F}}(t)$ for all $t\in\Real$. Therefore, for each realization $\widetilde{\tau}$ will belong to the set of FDR-controlling thresholds with probability 1, i.e., $\widetilde{\tau}\in\bigcup_{0<\alpha'\leq \alpha}\big\{t\geq 0:\Fc(t)=t/\alpha'\big\}$ a.s.
% \end{proof}
% \begin{remark}
% From a probabilistic point of view, $\widetilde{\mathcal{F}}(t)\leq {\mathcal{F}}(t)$ for all $t\in\Real$ is known as $\widetilde{\mathcal{F}}(t)$ stochastically dominates ${\mathcal{F}}(t)$.
% \end{remark}

% \begin{fact}
% $\hat{\mathcal{F}}(t)\leq {\mathcal{F}}(t)$ for all $t\in\Real$.
% \end{fact}
% \begin{proof}

% \end{proof}

\begin{propo}
Our sample-and-forward algorithm controls the FDR globally.\label{p1}
\end{propo}
\begin{IEEEproof}
Let $\hat{R}$ denote the total number of rejections in the network made according to $\hat{\tau}$. We note that $\hat{R}=m\mathcal{F}(\hat{\tau})\geq m\hat{\mathcal{F}}(\hat{\tau})= m{\hat{\tau}}/{\alpha}$ by $\hat{\mathcal{F}}\leq {\mathcal{F}}$ and \eqref{eq:BH-th}. Therefore,
\begin{align*}
    \text{FDR} = \sum_{j\in\mathcal{H}_0}\mathbb{E}\Bigg[\frac{\ind\big\{P_j\leq \hat{\tau}\big\}}{\hat{R}\vee 1}\Bigg]
    &\leq\sum_{j\in\mathcal{H}_0}\mathbb{E}\Bigg[\frac{\ind\big\{P_j\leq \hat{\tau}\big\}}{(m/\alpha)\hat{\tau}\vee 1}\Bigg],
    % &= \sum_{x\in\mathcal{V}}\sum_{j\in\mathcal{H}_0^{(x)}}\mathbb{E}\Bigg[\Big(\frac{\alpha'_x}{m}\Big)\frac{\ind\big\{P_j^{(x)}\leq {\tau}^{(x)}\big\}}{{\tau}^{(x)}}\Bigg]\\
   % &= \sum_{j\in\mathcal{H}_0}\frac{\alpha}{m}\mathbb{E}\Bigg[\frac{\ind\big\{P_j\leq \hat{\tau}\big\}}{\hat{\tau}\vee 1}\Bigg] ,
\end{align*}
where $\Hc_0=\{1 \leq k\leq m:\mathsf{H_{0,k}}\text{ is true}\}$. We note that ${\hat{\tau}}$ is a non-increasing function of any p-value in the network since decreasing a p-value will result in a stochastically smaller sampled pooled pseudo-CDF $\hat{\mathcal{F}}$ and thus larger ${\hat{\tau}}$. Therefore, according to~\cite{blanchard2008two,ramdas2019unified}, we get $\mathbb{E}\left[\frac{\ind\big\{P_j\leq \hat{\tau}\big\}}{\hat{\tau}\vee 1}\right]\leq 1$.
% \begin{equation}
%     \mathbb{E}\Bigg[\frac{\ind\big\{P_j\leq \hat{\tau}\big\}}{\hat{\tau}\vee 1}\Bigg]\leq 1\,.
% \end{equation}
Hence, $\text{FDR} \leq \sum_{j\in\mathcal{H}_0}\frac{\alpha}{m}=\frac{m_0}{m}\alpha$, where $m_0$ denotes the total number of null p-values in the network.
\end{IEEEproof}
% \smallskip 

\subsection{Power Analysis}
\label{sec:quality}
 It is of interest to understand the approximation quality of $\hat{\mathcal{F}}(t)$ in~\eqref{eq:pooling} (under the fixed sampling location scheme with $t_j=\frac{j-1}{M-1}\a$). Observe that if $\tau_{\text{BH}}=0$, then $\hat\tau=0$ since $\hat\tau\leq \tau_{\text{BH}}$ by definition. On the other hand, if $\tau_{\text{BH}}>0$ and the p-values are generated according to a continuous distribution function, with probability 1, one can take the sample size $M$ large enough such that $p_{(i)}< t_j<\tau_{\text{BH}}<p_{(i+1)}$ for some $1\leq j\leq M$. In this case, we get $\hat{\mathcal{F}}(\tau_{\text{BH}})={\mathcal{F}}(\tau_{\text{BH}})$, which implies $\hat{\tau}=\tau_{\text{BH}}$, immediately. 
 To shed light on the approximation quality in terms of $M$, suppose the p-values are generated \iid according to the CDF $F=\pi_0 U+\pi_1 G,\ \pi_1>0$, where 
 %$U(t)=t,\ 0\leq t\leq 1$
 $U$ is the distribution function of true nulls and $G(t)$ denotes the common CDF of the p-values under true alternatives. Note that $G$ can be a mixture of various alternative distributions. Define $\tau^*:=\sup\left\{t\in\Real:F(t)=t/\alpha\right\}$.

 \begin{assumption}\label{a2}
$F(t)>t/\alpha,\ t\in(\tau^*-\delta,\tau^*)$ for some $\delta>0$, $F(t)$ is continuously differentiable at $\tau^*$, and $F'(\tau^*)\neq 1/\alpha$. 
% In this case, Corollary \ref{c1} and \ref{c2} hold for $M>(\alpha/\delta)\ind\{\tau^*>2\delta\}$ and $F'(\overline{\tau})$ in Corollary \ref{c2} needs to be replaced with the smallest Lipschitz constant for $F$ over the interval $[t_j,t_{j^*+1})$ that contains $\tau^*$. 
\end{assumption}
%  \begin{assumption}\label{a2}
% The equation $G(t)=t/\alpha$ admits a unique non-zero solution ($\tau^*>0$), $G(t)$ is differentiable at $\tau^*$, and $G'(\tau^*)\neq 1/\alpha$.
%  \end{assumption}

 %if $\beta := \frac{(1/\a)-{r}_0}{r_1}\neq F'(0)$, from \cite{genovese2002operating} we know that 
%  \begin{equation*}
%     \tau_{\text{BH}}\xrightarrow{{P}}\tau^*\quad \text{as}\ m\rightarrow\infty .
%     %\sup\{t\in\Real:F(t)=\beta\,t\}
% \end{equation*} 

% If Assumption \ref{a2} holds, from \cite{genovese2002operating} we know $\tau_{\text{BH}}\xrightarrow{P}\tau^*$ as $m\rightarrow\infty$. 
% The following lemma (see Appendix A for the proof) is a corollary of this result and it is proved in \cite{https://doi.org/10.1111/j.1467-9868.2004.00439.x} under a slightly different set of assumptions. 
 The following lemma is a stronger version of \cite[Theorem~1]{genovese2002operating}, which only relies on continuous differentiability of $F(t)$ at $\tau^*$, and $F'(\tau^*)\neq 1/\alpha$ (part of Assumption~\ref{a2}).

\smallskip
\begin{lemma}(\!\!~\cite[Lemma~4]{pournaderi2022large})\label{l2}
If Assumption \ref{a2} holds, then $\tau_{\text{BH}}\xrightarrow{a.s.}
    \tau^*$.
%  \begin{equation*}
%     \tau_{\text{BH}}\xrightarrow{a.s.}
%     \tau^*\quad \text{as}\ m\rightarrow\infty .
% \end{equation*} 
\end{lemma}
\smallskip

% \begin{proof}
% The proof is given in Appendix A.
% \end{proof}
% \begin{assumption}\label{a3}
% $G(t)$ is differentiable on $(0,\eps)$ for some $\eps > 0$ and $G'_+(0)\neq 1/\alpha$.
% \end{assumption}
% \begin{assumption}\label{a3}
% $G(t)>t/\alpha$ for $0<t<\tau^*$. Note that if $G$ is continuous, then this assumption holds automatically as a consequence of Assumption \ref{a2}.
% \end{assumption}
\begin{assumption} \label{a4}
$t_{j^*}< \tau^* < t_{j^*+1}$ for some $1\leq j^*\leq M$, meaning, $\tau^*$ is not located exactly at some sampling point $t_j$.
\end{assumption}

The following theorem (see Appendix B for the proof) concerns the limiting value of the sample-and-forward threshold.

\smallskip
\begin{theorem}\label{t1}
If Assumptions \ref{a2} (with $\delta=\delta^*$) and \ref{a4} hold, then $\hat\tau\xrightarrow{a.s.}
    \alpha\,F(t_{j^*})=:\overline{\tau}$
% \begin{equation*}
%     \hat\tau\xrightarrow{a.s.}
%     \alpha\,G(t_j)=:\overline{\tau}\quad \text{as}\ m\rightarrow\infty 
% \end{equation*}
for all $M>(\alpha/\delta^*)\ind\{\tau^*>2\delta^*\}+1$.
\end{theorem}
\smallskip

% \begin{proof}
% The proof is given in Appendix B.
% \end{proof}
\begin{corollary}\label{c1}
If Assumptions \ref{a2} (with $\delta=\delta^*$) and \ref{a4} hold, then $\underset{m\rightarrow\infty}{\lim} \hat{\tau}\in(t_{j^*},t_{j^*+1}),\ a.s.$ and therefore, $\underset{m\rightarrow\infty}{\lim} \left(\tau_{\text{BH}}-\hat{\tau}\right)\leq\alpha/(M-1),\ a.s.$
% \begin{equation*}
%     \underset{m\rightarrow\infty}{\lim} \left(\tau_{\text{BH}}-\hat{\tau}\right)\leq\alpha/M\qquad \text{a.s.}
% \end{equation*}
for all $M>(\alpha/\delta^*)\ind\{\tau^*>2\delta^*\}+1$.
\end{corollary}
\begin{IEEEproof}
%We note that $\tau^*$ must satisfy $t_j\leq \tau^* < t_{j^*+1}$ for some $0\leq j\leq M$. 
% Since $t_{j^*+1}>\tau^*$, we get $G(t_{j^*+1})<t_{j^*+1}/\alpha$ according to the definition of $\tau^*$ and the fact that $G\leq 1$. From $\tau^*>t_{j^*}$ we get $G(t_{j^*})>t_{j^*}/\alpha$ since $\tau^*$ is the only non-zero solution to $G(t)=\alpha/t$ and $G'(\tau^*)\neq 1/\alpha$. 
% %If $\tau^*=t_{j^*}$, we get $G(t_{j^*}) = t_{j^*}/\alpha$ by the definition of $\tau^*$. 
% Putting the pieces together we get, $t_{j^*}/\alpha< G(t_{j^*})< G(t_{j^*+1}) < t_{j^*+1}/\alpha$.
% % \begin{equation*}
% %     t_{j^*}/\alpha< G(t_{j^*})< G(t_{j^*+1}) < t_{j^*+1}/\alpha\ ,
% % \end{equation*}
% which implies
According to the proof of Theorem \ref{t1} (\eqref{tau_hat_interval} and \eqref{slln}), we have $t_{j^*}< \alpha\,F(t_{j^*})=\overline{\tau} < t_{j^*+1}$ for $M>(\alpha/\delta^*)\ind\{\tau^*>2\delta^*\}+1$. We note that $\hat\tau\leq \tau_{\text{BH}}$ (because $\hat{\mathcal{F}}\leq {\mathcal{F}}$). Therefore, $t_{j^*}<\overline{\tau}\leq \tau^* < t_{j^*+1}.$
% \begin{equation*}
%     t_{j^*}\leq \underset{m\rightarrow\infty}{\lim} \hat{\tau}=\overline{\tau}\leq \underset{m\rightarrow\infty}{\lim} \tau_{\text{BH}}=\tau^* < t_{j^*+1}.
% \end{equation*}
%The result follows from $t_{j^*+1}-t_{j^*}=\alpha/M$.
\end{IEEEproof}
\begin{assumption}\label{a5}
$G$ is $C$-Lipschitz on $(t_{j^*},t_{j^*+1})$.
\end{assumption}
% We note that Assumption \ref{a5} is a stronger version of Assumptions~\ref{a2}+\ref{a3}.
Let $\Pc_{\text{BH}}(m)$ and $\hat\Pc(m)$ denote the detection power of the BH and sample-and-forward procedures. See Appendix C for the proof of the following corollary.
\begin{corollary}\label{c2}
If Assumptions \ref{a2} (with $\delta=\delta^*$), \ref{a4} and \ref{a5} (with $C=C^*$) hold, then $\underset{m\rightarrow\infty}{\lim} \left(\Pc_{\text{BH}}(m)-\hat\Pc(m)\right)\leq C^*\,\frac{\alpha}{M-1} $
% \begin{equation}
%      \underset{m\rightarrow\infty}{\lim} \left(\Pc_{\text{BH}}(m)-\hat\Pc(m)\right)\leq C^*\,\alpha/M 
% \end{equation}
for $M>(\alpha/\delta^*)\ind\{\tau^*>2\delta^*\}+1$.
\end{corollary}
% \begin{proof}
% The proof is given in Appendix C.
% \end{proof}
% \begin{proof}
% According to the Glivenko-Cantelli Theorem~\cite{van2000asymptotic} and $\tau_{\text{BH}}\xrightarrow{{a.s.}}\tau^*$, we get $\text{TDP}(\tau_{\text{BH}})\xrightarrow{a.s.}G(\tau^*)$. Dominated convergence theorem implies,
% %the weak law of large numbers and boundedness of $\frac{R-V}{m_1\vee 1}\leq 1$ imply,
% \begin{align*}
%     \underset{m\rightarrow\infty}{\lim} \left(\Pc_{\text{BH}}-\hat\Pc\right)=G(\tau^*)-G(\overline{\tau})
%     &\leq C^*\big(\tau^*-\overline{\tau}\big)\\
%     &\leq C^*\,\alpha/(M-1) ,
% \end{align*}
% where the inequalities hold according to Corollary \ref{c1} and Assumption \ref{a5}.
% \end{proof}
\begin{remark}
Note that for Gaussian statistics, Assumption~\ref{a2} holds for $\delta=\tau^*$ and Assumption~\ref{a5} holds with $C=G'(t_{j^*}),\ \text{if } M\geq 3$. Therefore, Theorem \ref{t1} and Corollary \ref{c1}, hold for all $M\geq 2$ and Corollary \ref{c2} holds for all $M\geq 3$.
\end{remark}

\section{Multihop Network with General Topologies}
In this section, we present a communication-efficient algorithm for the general multihop networks, adapting our method to an existing algorithm for decentralized inference known as the QuTE \cite{Ramdas2017b}, which controls the FDR over networks with arbitrary topology by allowing communication between adjacent nodes. 
%This algorithm controls the FDR over networks with arbitrary topology by allowing communication between adjacent nodes (connected by an edge in the graph), improving the detection power upon the Bonferroni method without communication.
The QuTE algorithm consists of three steps: query, test, and exchange. At the query step, each node needs to communicate all of its p-values to the neighboring nodes. We now present our communication-efficient version of this algorithm where nodes communicate sampled pseudo-CDFs (instead of p-values) while maintaining the FDR control.
% \begin{itemize}\setlength\itemsep{0.5em}
% 	\item[(1)] {\bf Query}:  
%     Each node queries its neighbors for the p-values they possess. 
%  	\item[(2)] {\bf Test}: 
%     Let $n^{(a)}$ denote the number of p-values node ``a" has after the communications in step (1). In this step, each node ``x" performs the BH procedure with $\alpha^{(x)}=({n^{(x)}}/{m})\alpha$ on its (pooled) p-values and computes the rejection threshold $\tau^{(x)}_0$.
% 	\item[(3)] {\bf Exchange}: Each node ``x" communicates $\tau_0^{(x)}$ to its neighbors. Suppose node ``a" has $N_a$ neighbors and Let  $\{\tau_{1}^{(a)},\hdots,\tau_{N_a}^{(a)}\}$ denote the set of thresholds node ``a" receives from its neighbors. Each node ``x" decides about the p-values that originally belong to it based on $\tau^{(x)}=\underset{0\leq i\leq N_x}{\max}\tau_{i}^{(x)}$, i.e., it rejects $\{1\leq j\leq m^{(x)}:P_j^{(x)}\leq \tau^{(x)}\}$. Note that if $i\neq 0$, then $\tau_i^{(x)}=\tau_0^{(y)}$ for some ``y" that neighbors ``x". 
% \end{itemize}

  \begin{figure*}[t!]\centering
\includegraphics[scale=0.33]{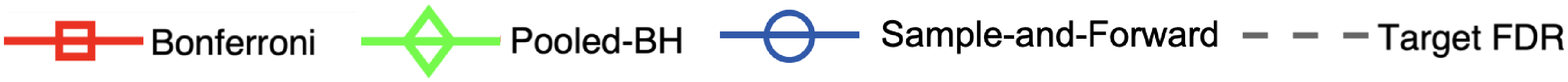}
\vspace{-1em}
\end{figure*}
\begin{figure*}[h]\centering
\begin{subfigure}{}\includegraphics[scale=0.33]{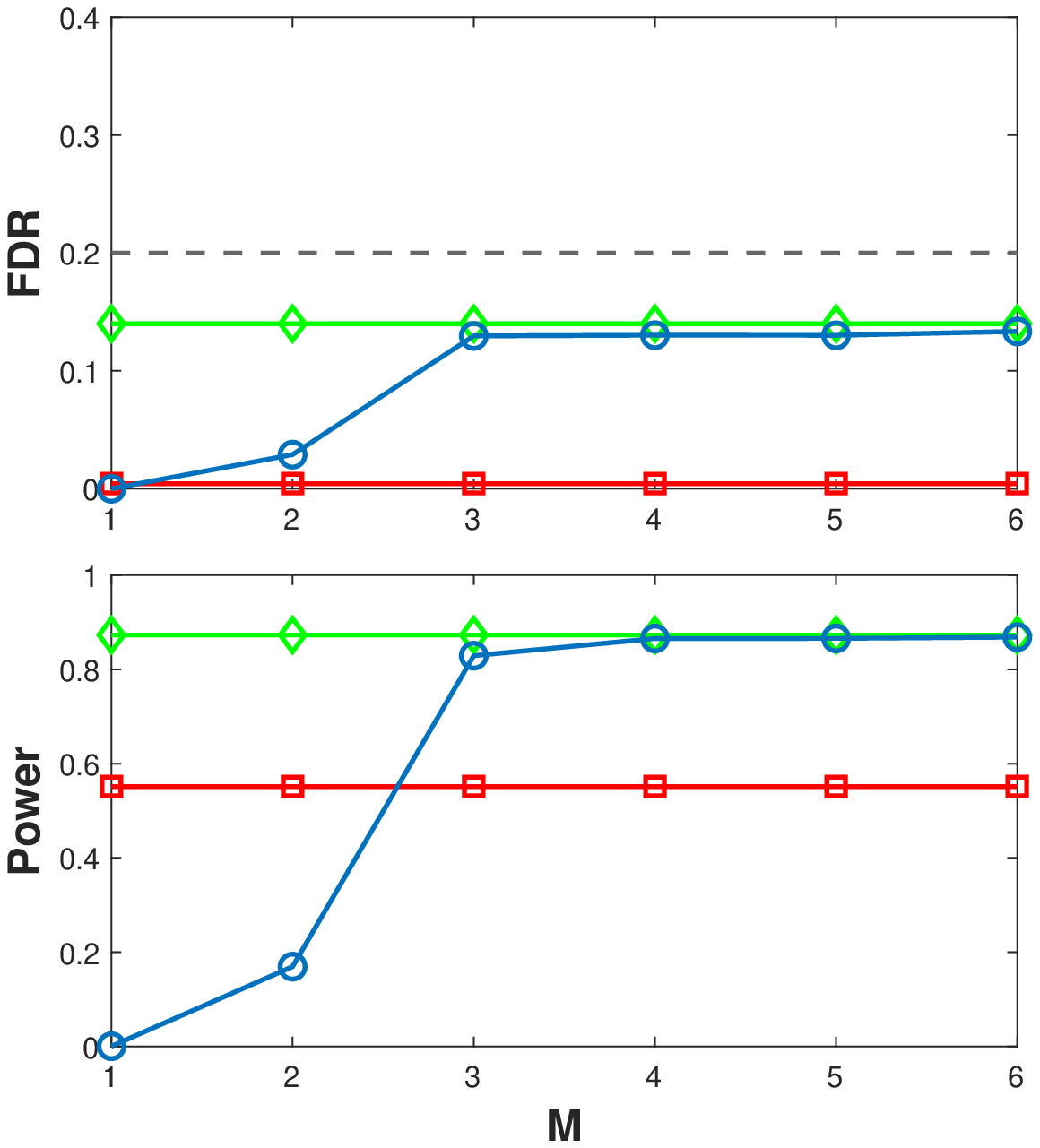}\end{subfigure}
\begin{subfigure}{}\includegraphics[scale=0.33]{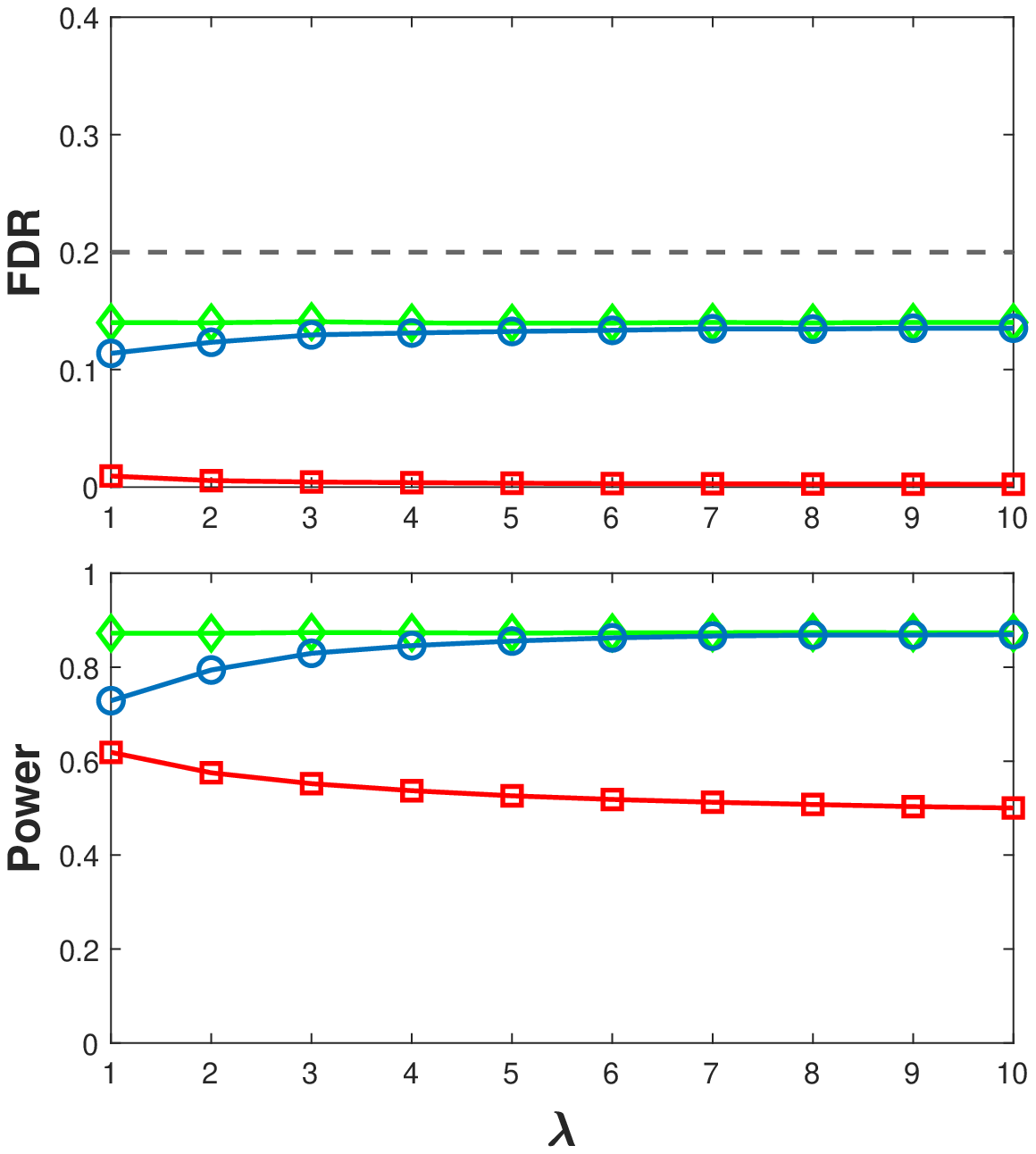}\end{subfigure}
\begin{subfigure}{}\includegraphics[scale=0.33]{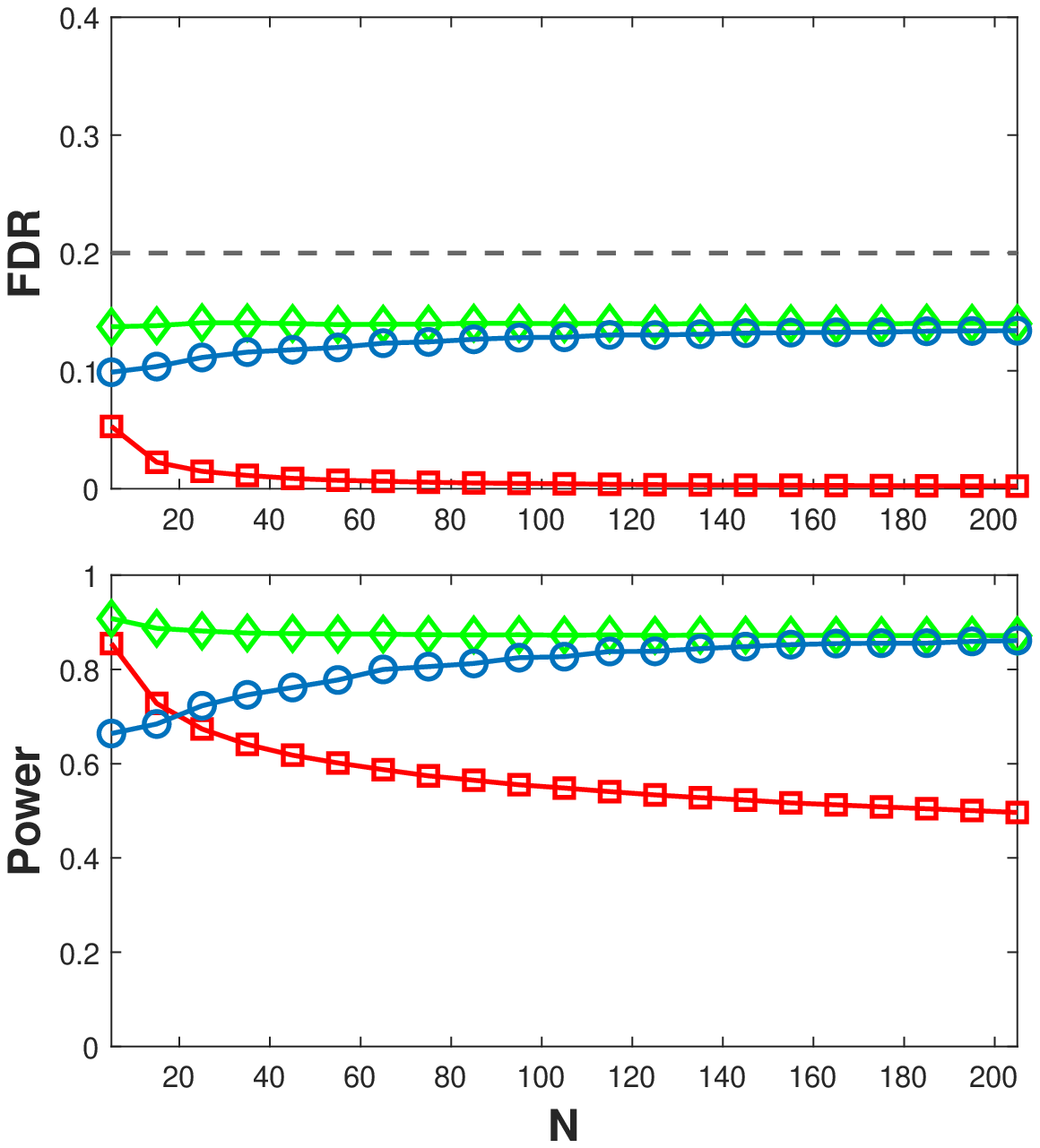}\end{subfigure}
\begin{subfigure}{}\includegraphics[scale=0.33]{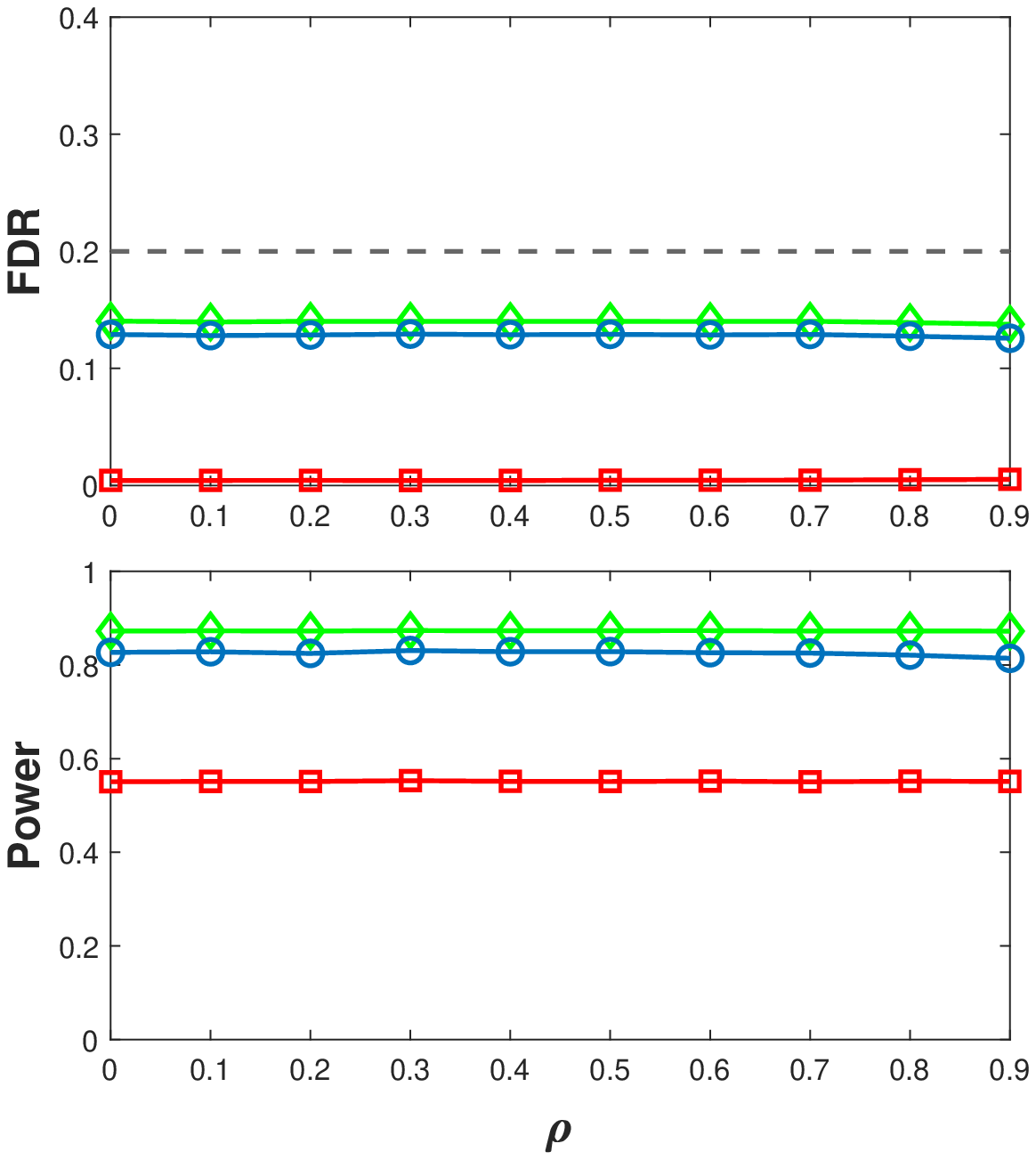}\end{subfigure}
\caption{ From left to right, the four sets of plots correspond to Experiments $1$ to $4$. $M$ denotes the number of samples taken at each node, $\l$ the mean of the number of p-values at each node, $N$ the number of nodes, and $\rho$ the correlation coefficient.}
\vspace{-1em}
\end{figure*}

\begin{itemize}\setlength\itemsep{0.1em}
	\item[(1)] {\bf Query}:  
    Each node queries its neighbors for the sampled pseudo-CDFs and the number of p-values they possess. 
 	\item[(2)] {\bf Test}: 
    Let $n^{(a)}$ denote the number of p-values at node $a$ and all of its neighbors. In this step, each node $x$ pools the sampled pseudo-CDFs using \eqref{eq:pooling} and computes the threshold $\hat{\tau}^{(x)}_0$ via \eqref{eq:BH-th} with the test size $\alpha^{(x)}=({n^{(x)}}/{m})\alpha$. 
	\item[(3)] {\bf Exchange}: Each node $x$ communicates $\hat{\tau}_0^{(x)}$ to its neighbors. Let $\{\hat{\tau}_{1}^{(a)},\hdots,\hat{\tau}_{N_a}^{(a)}\}$ denote the set of thresholds node $a$ receives from its neighbors where $N_a$ denote the number of the nodes neighboring node $a$. Each node $x$ decides about the p-values that originally belong to it based on $\hat{\tau}^{(x)}=\underset{0\leq i\leq N_x}{\max}\hat{\tau}_{i}^{(x)}$, i.e., it rejects $R_{(x)}=\{1\leq j\leq m^{(x)}:P_j^{(x)}\leq \hat{\tau}^{(x)}\}$. Note that if $i\neq 0$, then $\hat{\tau}_i^{(x)}=\hat{\tau}_0^{(y)}$ for some $y$ that neighbors $x$. 
\end{itemize}

%  To make this algorithm communication-efficient, we propose to communicate the sub-sampled pseudo-CDF's instead of p-values as described in \ref{sec:alg}. More precisely, in the query step each node ``x" take samples from the pseudo-CDF $\mathcal{F}^{(x)}(t)$ and share it with its neighbors along with (the number of p-values it has) $m^{(x)}$, just as step (1) in our algorithm. In the test stage of the algorithm, each node ``x" pools the sub-sampled pseudo-CDf's using \eqref{eq:pooling} and compute the threshold $\hat{\tau}^{(x)}_0$ via \eqref{eq:BH-th} and with the test size $\alpha^{(x)}=({n^{(x)}}/{m})\alpha$. The exchange step remains the same, i.e., $\hat{\tau}^{(x)}=\underset{0\leq i\leq N_x}{\max}\hat{\tau}_{i}^{(x)}$.

 \begin{propo}
 Our sample-and-forward version of the QuTE algorithm controls the FDR globally.
 \end{propo}

\begin{IEEEproof}
We follow the approach in \cite{Ramdas2017b} closely. Let $\mathcal{V}$ and $\mathcal{H}_0^{(a)}$ denote the set of nodes and the set of null p-values at node $a$, respectively. For node $a$, define $\breve{R}_0^{(a)}=n^{(a)}\frac{\hat{\tau}_0^{(a)}}{\alpha^{(a)}}$ and let $\{\breve{R}_{1}^{(a)},\hdots,\breve{R}_{N_a}^{(a)}\}$ denote the same quantity for its neighbors. Now define $\breve{R}^{(a)}=\underset{0\leq i\leq N_a}{\max}\breve{R}_{i}^{(a)}$ and observe $\breve{R}^{(a)}=m\frac{\hat{\tau}^{(a)}}{\alpha}$. Let $\hat{R}$ denote the total number of rejections in the network using the communication-efficient QuTE algorithm. 
%Let $\hat{R}$ and $\hat{R}^{(a)}$ denote the total number of rejections in the network and maximum number of rejections over node $a$ and its neighbors using the communication-efficient QuTE algorithm. Note that $\hat{R}^{(a)}$ is different from $R_{(a)}$ defined at step (3) of the algorithm in the sense that $R_{(a)}$ only counts the rejected p-values (according to maximum threshold $\hat{\tau}^{(a)}$) that originally belong to $a$ while $\hat{R}^{(a)}$ takes the maximum over the number of rejections where the number of rejections are calculated based on the total number of p-values 
We have
\begin{align*}
    \text{FDR} &= \sum_{x\in\mathcal{V}}\sum_{j\in\mathcal{H}_0^{(x)}}\mathbb{E}\Bigg[\frac{\ind\big\{P_j^{(x)}\leq \hat{\tau}^{(x)}\big\}}{\hat{R}\vee 1}\Bigg]=:\sum_{x\in\mathcal{V}}\sum_{j\in\mathcal{H}_0^{(x)}}A_{x,j},
\end{align*}
where $A_{x,j}$ can be upper bounded as follows.
\begin{align*}
   A_{x,j}\leq\mathbb{E}\Bigg[\frac{\ind\big\{P_j^{(x)}\leq \hat{\tau}^{(x)}\big\}}{\breve{R}^{(x)}\vee 1}\Bigg]
     =\mathbb{E}\Bigg[\frac{\ind\big\{P_j^{(x)}\leq \hat{\tau}^{(x)}\big\}}{(m/\alpha)\,\hat{\tau}^{(x)}\vee 1}\Bigg],
    % &= \sum_{x\in\mathcal{V}}\sum_{j\in\mathcal{H}_0^{(x)}}\mathbb{E}\Bigg[\Big(\frac{\alpha'_x}{m}\Big)\frac{\ind\big\{P_j^{(x)}\leq {\tau}^{(x)}\big\}}{{\tau}^{(x)}}\Bigg]\\
%    &= \frac{\alpha}{m}\,\mathbb{E}\Bigg[\frac{\ind\big\{P_j^{(x)}\leq {\hat{\tau}}^{(x)}\big\}}{{\hat{\tau}}^{(x)}\vee 1}\Bigg].
\end{align*}
 because $\hat{R}\geq\hat{R}^{(y)}\geq\breve{R}^{(y)}_0=\breve{R}^{(x)}$ for some $y\in x\cup\text{neighbors}(x)$, where $\hat{R}^{(y)}$ denotes the number of p-values at node $y$ and its neighbors that are smaller than $\hat\tau^{(y)}$. By the same argument as in the proof of Proposition \ref{p1}, ${\hat{\tau}}^{(x)}$ is a non-increasing function of any p-value at node $x$ or at its neighbors.
 %since decreasing a p-value can only result in a stochastically smaller pseudo-CDF which implies stochastically smaller sampled pseudo-CDF and larger ${\hat{\tau}}^{(x)}$. 
 Thus, according to~\cite{blanchard2008two,ramdas2019unified}, we get $\mathbb{E}\left[\frac{\ind\big\{P_j^{(x)}\leq {\hat\tau}^{(x)}\big\}}{{\hat\tau}^{(x)}\vee 1}\right]\leq 1$.
% \begin{equation}
%     \mathbb{E}\Bigg[\frac{\ind\big\{P_j^{(x)}\leq {\hat\tau}^{(x)}\big\}}{{\hat\tau}^{(x)}\vee 1}\Bigg]\leq 1\,.
% \end{equation}
Hence, $\text{FDR} \leq \sum_{x\in\mathcal{V}}\sum_{j\in\mathcal{H}_0^{(x)}}\frac{\alpha}{m}=\frac{m_0}{m}\alpha$, where $m_0$ denotes the total number of null p-values in the network.
\end{IEEEproof}

%\begin{remark}
The QuTE algorithm has been extended to the setting where each node can query from the neighbors that are $c$ edges away, which requires $c$ rounds of communication between immediate neighbors~\cite{Ramdas2017b}. Our arguments regarding the communication efficiency and FDR control via sampling the pseudo-CDFs can be carried over to $c\geq 2$ as well. Note that the ``Query" step plays the role of synchronization in the multi-hop setting. 
%\end{remark}

\section{Numerical Results}
In this section, we evaluate the empirical performance of our algorithm for the star networks with $\alpha = 0.2$. The \emph{estimated} FDR and power are computed by averaging over $10000$ trials. 
The number of p-values at each local node is drawn from $\Poisson(\lambda)$ and the probability of generating a non-null p-value at node $i$ is $\pi_1^{(i)}=0.5-0.4(i/N)$, where $N$ denotes the number of local nodes in the network. The statistics are distributed according to $\mathcal{N}(0,1)$ under $\mathsf{H_0}$ and we compute two-sided p-values.
% The p-values are generated according to $\mathcal{U}[0,1]$ under $\mathsf{H_0}$.
Under $\mathsf{H_1}$, 
%we consider both simple alternatives ($\mathcal{N}(\mu,1)$ for $\mu\ne 0$) and mixture alternatives (first picking $\mu\in [\mu_{\min}, \mu_{\max}]$ and then generating a sample according to $\mathcal{N}(\mu,1)$). 
we consider 
%mixture alternatives (i.e., $\mathcal{N}(\mu,1)$ with random $\mu$ and we fix the distribution of $\mu$ for all nodes) and 
composite alternatives, i.e., we generate samples according to $\mathcal{N}(\mu^{(i)},1)$ at node $i$ where $\mu^{(i)}\sim\U\{ [\mu_{\text{base}}^{(i)}-0.5, \mu_{\text{base}}^{(i)}+0.5]\}$ and $\mu_{\text{base}}^{(i)}=2+4(i/N)$. The pseudo-CDFs at local nodes are sampled at $\frac{j}{M+1},\ 1\leq j\leq M$. The sample locations in section \ref{sec:alg} are chosen slightly differently than this just to simplify the presentation of our power analysis.
%with a unique distribution function for $\mu^{(i)}$). 
%We refer to our method as \emph{Sampled-BH} and compare it 
We compare our method sample-and-forward, with the two baselines discussed in Section~\ref{sec:quantized-FDR}: the  Bonferroni method and the global (referred to as \emph{pooled-BH}) multiple testing by carrying out the BH procedure over all the p-values from all nodes, i.e., $\{\Pv^{(1)},..., \Pv^{(N)}\}$.
Under this setting, we consider the following four experiments.
%Pooled-BH requires the nodes to communicate their entire set of p-values they own. However, the equivalent performance can be achieved by sending only the p-values less than $\alpha$ as well as the number of p-values each node possesses.
%To test the robustness of our method, we evaluate two settings in Experiment 4 and 5 that violate our assumptions for theoretical guarantees. 
% \begin{figure}[!h]
% \centering
%   \includegraphics[scale=0.5]{M_1_25.eps}
%   \caption{Experiment~1 (vary M). Number of nodes = 100.}\vspace{-1em}
% \end{figure}

%\smallskip
\noindent{{\bf Experiment 1 (vary M).} We examine the effect of varying $M$ on the power and FDR. We fix $N=100$ and $\lambda=3$. Our sample-and-forward method approaches the pooled BH even when the number of samples $M$ is as low as $3$.
%The probability of generating a non-null p-value at node $i$ is $\pi_1=0.5-0.4(i/N)$. We fix $\mu_{\text{base}}^{(i)}=2+4(i/N)$, and in each trial, generate samples under $\mathsf{H_1}$ in node $i$ by first picking $\mu^{(i)}\sim\U\{ [\mu_{\text{base}}^{(i)}-0.5, \mu_{\text{base}}^{(i)}+0.5]\}$
%$\mu\sim\U\{[-\mu_{\text{base}}^{(i)}-0.5, -\mu_{\text{base}}^{(i)}+0.5]\cup [\mu_{\text{base}}^{(i)}-0.5, \mu_{\text{base}}^{(i)}+0.5]\}$
%and then generating $X^{(i)}_j\sim \mathcal{N}(\mu^{(i)},1)$. 
%It can be observed that a phase transition occurs at sample size $M=5$.
}

% \begin{figure}[h]
% \centering
%   \includegraphics[scale=0.4]{M_1_100.eps}
%   \caption{Experiment~1 ($M$ denotes the number of samples taken from the pseudo-CDFs.)}\vspace{-1em}
% \end{figure}

%\smallskip
\noindent{{\bf Experiment 2 (vary $\lambda$).} We fix $M=3,\ N=100$ and vary $\lambda$ from $1$ to $10$. 
%The setting is otherwise the same as in Experiment~1.
We observe that 
for a fixed number of samples taken from the pseudo-CDFs, 
increasing the number of p-values improves the power of the sample-and-forward while it impacts the Bonferroni method in a negative way.}

% \begin{figure}[h]
% \centering
%   \includegraphics[scale=0.4]{m_nnodes20.eps}
%   \caption{Experiment~2 ($n$ denotes the number of p-values at each node.)}\vspace{-1em}
% \end{figure}

% \smallskip
% \noindent{{\bf Experiment 3 (Vary $\mu$).} In this experiment, we fix $\mu_{\text{base}}^{(i)}=\mu$ and vary $\mu$ from $2$ to $4$. The number of samples taken from the pseudo-CDFs is set to $M=5$. The setting is otherwise the same as in Experiment~1.}

% \begin{figure}[h]
% \centering
%   \includegraphics[scale=0.4]{mu_nnodes100_m1_100.eps}
%   \caption{Experiment~3 (Vary $\mu$)}\vspace{-1em}
% \end{figure}

%\smallskip
\noindent{{\bf Experiment 3 (vary N).} We fix $M=3$ and $\lambda=3$, and vary the number of nodes in the network from $N=5$ to $205$. 
%The p-values are generated in the same way as in Experiment~1.
It can be observed that increasing the number of nodes improves the power of the sample-and-forward while it impacts the Bonferroni method in a negative way.}

% \begin{figure}[h]
% \centering
%   \includegraphics[scale=0.4]{n_nodes_n25_mu3_a025.eps}
%   \caption{Experiment~4 ($N$ denotes the number of nodes.)}\vspace{-1em}
% \end{figure}

%\smallskip

% \noindent{{\bf Experiment 5 (Dependent p-values).} Finally, we evaluate the robustness of our method by considering dependent p-values. In particular, we adopt two commonly used covariance structures: (I) $\Sigma_{i,j} = \rho^{|i-j|}$, and (II) $\Sigma_{i,i} = 1$, $\Sigma_{i,j} = \rho\cdot\ind (\lceil i/10\rceil=\lceil j/10\rceil)$, where $\ind(\cdot)$ denotes the indicator function. We vary $\rho$ from $0$ to $0.8$. We have $N=30$ nodes and $M=5$. The rest of the settings is the same as in Experiment~1.}

\noindent{{\bf Experiment 4 (dependent p-values).} We fix $M=3,\ \lambda=3$, and $N=100$, and evaluate the performance of our method when the p-values are dependent. In particular, we adopt the tapering covariance structure $\Sigma_{i,j} = \rho^{|i-j|}$ where $\rho$ varies from $0$ to $0.9$. Although these p-values are not known to satisfy Assumption $\ref{prds}$, it can be observed that the sample-and-forward has stable power and controls the FDR in this setting.}

% \section{Conclusion}
% In this paper, we have shown that in a network with arbitrary topology and total of $m$ p-values, communicating $\mathcal{O}(\log m)$ bits by each node can result in a detection power close to the centralized performance where each node communicates $\mathcal{O}(m)$ (real-valued) p-values to its neighboring nodes. We have proved that our method, sample-and-forward, preserves the rigorous FDR control.

% \begin{figure}[h]
% \centering
%   \includegraphics[scale=0.4]{rho_nnodes30_m1_20.eps}
%   \caption{Experiment~5}\vspace{-1em}
% \end{figure}

%\balance
% \begin{figure}[h]
% \centering
%   \includegraphics[scale=0.5]{rho_II_nnodes30_m1_30.eps}
%   \caption{Experiment~5 with covariance structure (II).}\vspace{-1.3em}
% \end{figure}

% \section{Conclusion}
% We have developed a simple yet effective algorithm for communication-efficient multiple testing problem in networks. The method has finite-sample provable FDR control and only requires transmitting $\mathcal{O}(\log m)$ bits, improving upon the existing methods. The detection power is analyzed under a mixture model, showing that a moderate number of samples guarantees good power, which is supported by our empirical results.

%This is supported by our empirical results, where our method shows competitive performance and robustness even with a small number of samples. 

\section{Conclusion}
We have proposed the sample-and-forward method, which communicates samples of pseudo-CDFs as a communication-efficient alternative for the transmission of (quantized) p-values to perform the BH (or more generally QuTE algorithm) in networks. Our method comes with the finite-sample FDR control, while the power loss has been characterized asymptotically. The numerical results confirm that a node receiving a constant number of pseudo-CDF samples from its neighbors with $\mathcal{O}(\log m)$ bits performs similarly to the centralized procedure with $\mathcal{O}(m\log m)$ bits.

% To prove the extended version (discussed in Remark \ref{r3}) one can make the same argument but with $\tau^*-\delta<\zeta<\tau^*$. The condition $M>(\alpha/\delta)\ind\{\tau^*>2\delta\}$ ensures
% \begin{equation*}
%     \tau^*-\delta <t_j < \tau^* <t_{j^*+1}
% \end{equation*}
% under which the result follows. 
% \end{proof}

%%
%% If several appendices are needed, then the command
%%
% \appendices
%%
%% in combination with further \section-commands can be used.
%%%%%%

%\section*{Acknowledgment}
%
%We are indebted to Michael Shell for maintaining and improving
%\texttt{IEEEtran.cls}. 

%%%%%%
%% To balance the columns at the last page of the paper use this
%% command:
%%
% \balance
% \enlargethispage{-4cm} 
%%
%% If the balancing should occur in the middle of the references, use
%% the following trigger:
%%
% \IEEEtriggeratref{18}
%%
%% which triggers a \newpage (i.e., new column) just before the given
%% reference number. Note that you need to adapt this if you modify
%% the paper.  The "triggered" command can be changed if desired:
%\IEEEtriggercmd{\enlargethispage{-10cm}}
%%
%%%%%%

%%%%%%
%% References:
%% We recommend the usage of BibTeX:
%%

\appendices
\section{Proofs}
% \subsection{Proof of Lemma \ref{l2}}
% \begin{IEEEproof}
%  %Let $\eps_m=1/\log{m}$. 
%  Define the sequences $a_m = \frac{m\tau^*}{\alpha}(1-1/\log{m})\quad\text{and}$ and $b_m = \frac{m\tau^*}{\alpha}(1+1/\log{m})$.
% % \begin{equation*}
% %     a_m = \frac{m\tau^*}{\alpha}(1-1/\log{m})\quad\text{and}\quad b_m = \frac{m\tau^*}{\alpha}(1+1/\log{m})\ .
% % \end{equation*}
% %Let $R_{\text{BH}} = m/\alpha\, \tau_{\text{BH}}$ denote the BH deciding index. 
% In the proof of Theorem 1 from \cite{genovese2002operating}, it is shown that
% \footnote{The result is derived under the strict concavity assumption on $G(t),\ t\in \Real_+$. However, this assumption can be weakened as pointed out in~\cite{genovese2002operating}.} 
% \begin{equation*}
%     \mathbb{P}(R_{\text{BH}} > b_m) 
%     \leq m\, e^{-c_1{m}/{(\log{m})^2}}\rightarrow 0\ ,
% \end{equation*}
% \begin{equation*}
%     \mathbb{P}(R_{\text{BH}} < a_m) 
%     \leq  e^{-c_2{m}/{(\log{m})^2}}\rightarrow 0\ ,
% \end{equation*}
% for some constants $c_1>0,\ c_2>0$. But the RHS of both inequalities is summable in $m$. Therefore, according to the Borel-Cantelli lemma $ a_m \leq R_{\text{BH}}\leq b_m,\ a.s.$
% % \begin{equation*}
% %     a_m \leq R_{\text{BH}}\leq b_m\qquad a.s.
% % \end{equation*}
% for $m>\breve{m}(\omega)$, completing the proof.
% \end{IEEEproof}
\subsection{Proof of Theorem \ref{t1}}
% We prove the theorem under the Assumptions \ref{a2}, \ref{a3}, and \ref{a4}. 
%that the equation $G(t)=\alpha/t$ admits a unique non-zero solution $\tau^*$ where $G'(\tau^*)\neq 1/\alpha$ and $G'_+(0)\neq 1/\alpha$. Note that the unique nonzero solution together with $G'(\tau^*)\neq 1/\alpha$ and $G'_+(0)\neq 1/\alpha$ implies $G'(\tau^*)<1/\alpha$ and $G'_+(0)> 1/\alpha$. Under this setting we only deal with a simple (yet not trivial) set of solutions $\{t:G(t)=t/\alpha\}=\{0,\tau^*\}$.
% Under this setting
% To prove the extended version (discussed in Remark \ref{r3}) one needs to make the same argument for the two largest solutions instead of $\{0,\tau^*\}$.
% \begin{proof}
 Define the sequences $a_m = \frac{m\tau^*}{\alpha}(1-m^{-1/4})$ and $b_m = \frac{m\tau^*}{\alpha}(1+m^{-1/4})$. Recall that $R_{\text{BH}} = m/\alpha\, \tau_{\text{BH}}$ denotes the BH deciding index and note that $ a_m \leq R_{\text{BH}}\leq b_m,\ a.s.$ for large $m$ according to the proof of~\cite[Lemma~4]{pournaderi2022large}.
% \begin{equation*}
%     a_m = \frac{m\tau^*}{\alpha}(1-1/\log{m})\quad\text{and}\quad b_m = \frac{m\tau^*}{\alpha}(1+1/\log{m})\ .
% \end{equation*}
Pick some $\tau^*-\delta^*<\zeta<\tau^*$ and define,
\begin{align*}
    D_{\max}(m):=& \frac{m}{\alpha}\sup\big\{t: \Fc(t)\geq t/\alpha\big\}\\
    =& \max\bigg\{\{0\}\cup\Big\{1\leq k\leq m: P_{(k)}\le \frac{\alpha k}{m}\Big\}\bigg\},
\end{align*}
\begin{align*}
    D_{\min}(m;\zeta)&:= \frac{m}{\alpha}\inf\big\{t>\zeta: \Fc(t)\leq t/\alpha\big\}\\
    =&\min\bigg\{\{m\}\cup\Big\{\zeta_m < k < m: P_{(k+1)}>\frac{\alpha k}{m}\Big\}\bigg\}.
\end{align*}
 where $\zeta_m= {m\,\zeta}/{\alpha}$. We note that
\begin{equation*}
    D_{\max}=R_{\text{BH}}=\frac{m}{\alpha}\sup\big\{t: \Fc(t)= t/\alpha\big\}.
\end{equation*}
Therefore, by Lemma \ref{l2} we have $a_m \leq D_{\max}\leq b_m$ for large $m$ almost surely. This implies $\{t>\zeta: \Fc(t)= t/\alpha\big\}\neq\varnothing\ a.s.$
% \begin{equation*}
%     \{t>\zeta: \Fc(t)= t/\alpha\big\}\neq\varnothing\quad a.s.
% \end{equation*}
for large $m$. Thus, for large enough $m$ we have $D_{\min}\leq D_{\max}$ almost surely. Now we wish to prove $a_m \leq D_{\min}\leq b_m$ for large $m$ almost surely. First we note that, $D_{\max}\leq b_m\ a.s.$ implies $D_{\min}\leq b_m\ a.s.$ since $D_{\min}\leq D_{\max}\ a.s.$ for large $m$. So we only need to show $ D_{\min}\geq a_m\ a.s.$. We note,
\begin{align*}
   & \hspace{-0.6em}\mathbb{P}(D_{\min} < a_m) 
    \leq \mathbb{P}\Bigg(\underset{\zeta_m < k\,<\,a_m}{\bigcup}\big\{P_{(k+1)} > \alpha k/m\big\}\Bigg) \\
    &\leq \mathbb{P}\left(\underset{\zeta_m < k\,<\,a_m}{\bigcup}\bigg\{\sum_{i=1}^m \ind\big\{P_i\leq \alpha k/m\big\}\leq k\bigg\}\right)\\
    &\leq \sum_{\zeta_m < k\,<\,a_m}\mathbb{P}\left[\mu(k)-\sum_{i=1}^m \ind\big\{P_i\leq \alpha k/m\big\}\geq \mu(k)-k\right],
    % &\leq \sum_{\zeta_m < k\,<\,a_m}\mathbb{P}\Bigg[\mu(k)-\bigg(\sum_{i=1}^m \ind\big\{P_i\leq \alpha k/m\big\}\bigg)\\
    % &\qquad\qquad\qquad\qquad\qquad\qquad\qquad\qquad\qquad\geq \mu(k)-k\Bigg],
\end{align*}
where $\mu(k) =  \mathbb{E}\left(\sum_{i=1}^m \ind\big\{P_i\leq \alpha k/m\big\}\right)= m F(\alpha k/m)$.
% \begin{align}
%   \mu(k) =  \mathbb{E}\left(\sum_{i=1}^m \ind\big\{P_i\leq \alpha k/m\big\}\right)= m F(\alpha k/m)\ .
%   %& =r_0\,m\,(k/m)\alpha + (1-r_0)\,m F\big(\alpha k/m\big)\ .\nonumber
% \end{align}
We observe that
\begin{align}\label{diff}
     \mu(k)-k = m \left(F(\alpha k/m)-\frac{1}{\alpha}(\alpha k/m)\right)=:h(k)\ .
\end{align}
According to Taylor's theorem we have,
\begin{equation*}
     h(a_m)= m\left({\tau^*}m^{-1/4}\big(1/\alpha-F'(\tau^*)\big)+o\big(m^{-1/4}\big)\right).
\end{equation*}
% Also,
% \begin{equation*}
%      h(\zeta_m)= {m}\big(G(\zeta)-\zeta/\alpha\big),
% \end{equation*}
% by the definition.

% Also, if $G'_+(0)<\infty$, we get 
% \begin{equation*}
%      h(\zeta_m)= m\Big(\frac{\alpha}{\log{\log{m}}}\big(G'_+(0)-1/\alpha\big)+o\big(\frac{1}{\log\log{m}}\big)\Big).
% \end{equation*}
% If $G'_+(0)=\infty$, we get
% \begin{equation*}
%      h(\zeta_m)> \frac{m\alpha}{\log{\log{m}}}\big(G'(\frac{\alpha}{\log{\log{m}}})-1/\alpha\big).
% \end{equation*}
We note,
\begin{align*}
    \underset{\zeta_m<k<a_m}{\inf}h(k)/m &=  \underset{\frac{\alpha \zeta_m}{m}<\frac{\alpha k}{m}<\frac{\alpha a_m}{m}}{\inf} \Big(F(\alpha k/m)-\frac{1}{\alpha}(\alpha k/m)\Big)\\
    &\geq \underset{\zeta<t<{\tau^*}(1-m^{-1/4})}{\inf} \Big(F(t)-\frac{t}{\alpha}\Big).
\end{align*}
According to Assumption~\ref{a2}, we have $F(t)-{t}/{\alpha}>0$ for all $\zeta<t<\tau^*$. But we note that,
\begin{equation*}
F\Big({\tau^*}\big(1-m^{-1/4}\big)\Big)-\frac{{\tau^*}\big(1-m^{-1/4}\big)}{\alpha}=\frac{h(a_m)}{m}=o(1),
\end{equation*}
and by Assumption~\ref{a2}, $F(t)-{t}/{\alpha}$ is strictly decreasing in a neighborhood of $\tau^*$.
Hence, for large enough $m$ we get 
$\underset{\zeta_m<k<a_m}{\inf}h(k)\geq h(a_m)$
% \begin{equation*}
%     \underset{\zeta_m<k<a_m}{\inf}h(k)/m\geq h(a_m)/m
% \end{equation*}
 which implies $h(k)\geq h(a_m)$ for all $\zeta_m< k<a_m$. 
 %We observe that $h(a_m)>0$ for large $m$ since $G'(\tau^*)<1/\alpha$ (from Assumption \ref{a2}) $G'(\tau^*)<1/\alpha$. 
 Now by Hoeffding's inequality, we get
\begin{align*}
    \mathbb{P}(D_{\min} < a_m)&
    \leq \sum_{\zeta_m < k\,<\,a_m}e^{-{2h(k)^2}/{m}}\\
    &\leq m\, e^{-{2h(a_m)^2}/{m}}\rightarrow 0\ . 
\end{align*}
Since the upper bound is summable in $m$, we get $D_{\min} \geq a_m\ a.s.$ for large $m$. Therefore, for any $\tau^*-\delta^*<\zeta<\tau^*$ we have
\begin{equation}\label{inf}
    \inf\{t > \zeta:\Fc(t)\leq t/\alpha\}\xrightarrow{a.s.}\tau^* .
\end{equation}
Also from Lemma \ref{l2} we have
\begin{equation}\label{sup}
    \sup\{t:\Fc(t)\geq t/\alpha\}\xrightarrow{a.s.}\tau^*.
\end{equation}
We note that if $M>(\alpha/\delta^*)\ind\{\tau^*>2\delta^*\}$ and $t_{j^*} < \tau^* <t_{j^*+1}$, then $\tau^*-\delta^* <t_{j^*} < \tau^* <t_{j^*+1}.$
In this case, according to \eqref{inf} and \eqref{sup} we get
\begin{equation*}
    t_{j^*}/\alpha < \Fc(t_{j^*}) <t_{j^*+1}/\alpha,\quad a.s.\ \ \text{all}\  m>m'(\omega),
\end{equation*}
which implies
\begin{equation}\label{tau_hat_interval}
    t_{j^*}< \alpha \Fc(t_{j^*}) <t_{j^*+1},\quad a.s.\ \ \text{all}\  m>m'(\omega).
\end{equation}
According to~\eqref{stime}, we get
\begin{align*}
    \hat{\Fc}\big(\alpha\Fc(t_{j^*})\big)&=\hat{\Fc}(t_{j^*})
    =\Fc(t_{j^*})=\frac{\alpha\Fc(t_{j^*})}{\alpha}.
\end{align*}
We note that $\hat\tau\leq \tau_{\text{BH}}$ (since $\hat{\mathcal{F}}\leq {\mathcal{F}}$) and therefore, $\hat{\tau}=\alpha\Fc(t_{j^*})$. By the strong law of large numbers, we have
\begin{equation}
    \Fc(t_{j^*})\xrightarrow{a.s.}\, F(t_{j^*})\label{slln}
\end{equation}
%$\Fc(t_{j^*})\xrightarrow{a.s.}\, G(t_{j^*})$
which implies $\hat\tau\xrightarrow{a.s.}\, \alpha F(t_{j^*})$, completing the proof.
\subsection{Proof of Corollary \ref{c2}}
\begin{IEEEproof}
According to the Glivenko-Cantelli Theorem~\cite{van2000asymptotic} and $\tau_{\text{BH}}\xrightarrow{{a.s.}}\tau^*$, we get $\text{TDP}(\tau_{\text{BH}})\xrightarrow{a.s.}G(\tau^*)$. Dominated convergence theorem implies,
%the weak law of large numbers and boundedness of $\frac{R-V}{m_1\vee 1}\leq 1$ imply,
\begin{align*}
    &\hspace{-0.3em}\underset{m\rightarrow\infty}{\lim} (\Pc_{\text{BH}}-\hat\Pc)=G(\tau^*)-G(\overline{\tau})\leq C^*\big(\tau^*-\overline{\tau}\big)
    \leq \frac{C^*\alpha}{M-1},
    %C^*\,\alpha/(M-1) ,
\end{align*}
where the inequalities hold according to Corollary~\ref{c1} and Assumption \ref{a5}.
\end{IEEEproof}

%%
%% where we here have assume the existence of the files
%% definitions.bib and bibliofile.bib.
%% BibTeX documentation can be obtained at:
%% http://www.ctan.org/tex-archive/biblio/bibtex/contrib/doc/
%%%%%%
\balance

 %\clearpage
\bibliographystyle{IEEEtran}
\bibliography{ref.bib}
\end{document}